\UseRawInputEncoding
\documentclass[letterpaper, 10 pt, conference]{ieeeconf}
\usepackage{etoolbox}
\newtoggle{IROS}
\settoggle{IROS}{true}

\iftoggle{IROS}
{ 
\IEEEoverridecommandlockouts    
\overrideIEEEmargins 
\usepackage{hyperref}
}{ 
\usepackage{times}
\usepackage[numbers]{natbib}
\usepackage{multicol}
\usepackage[bookmarks=true]{hyperref}
}

\usepackage{subcaption}
\usepackage[pdftex]{graphicx}
\usepackage{multirow}
\usepackage{multicol}

\usepackage{xcolor}

\usepackage{amsfonts}
\usepackage{amsmath}

\usepackage{amssymb} % more math

\usepackage{amsthm} % theorems
\usepackage{multirow}
\usepackage{makecell}
\usepackage{booktabs}
\usepackage{paralist}

\usepackage{dsfont}

\newcommand{\scenario}{\mathcal{S}}

\newcommand{\figref}[1]{Fig.\ref{#1}}

\newcommand{\highd}[0]{HighD~\cite{krajewski2018highd}}
\newcommand{\round}[0]{RounD~\cite{krajewski2020round}}
\newcommand{\sdd}[0]{SDD~\cite{robicquet2020learning}}
\newcommand{\vci}[0]{VCI-DUT~\cite{yang2019top}}
\newcommand{\ngsim}[0]{NGSIM~\cite{simulation2007us}}

\newtheorem{definition}{Definition}

\newtheorem{proposition}{Proposition}

\newcommand{\nitb}[1]{\par \noindent{\textbf{#1}}}

\title{Reactive and Safe Road User Simulations using Neural Barrier Certificates}

\iftoggle{IROS}
{
\author{Yue Meng$^{1}$, Zengyi Qin$^{1}$ and Chuchu Fan$^{1}$%
\thanks{$^{1}$Yue Meng, Zengyi Qin and Chuchu Fan are with the Department of Aeronautics and Astronautics, Massachusetts Institute of Technology, Cambridge, MA 02139, USA
{\tt\small \{mengyue,qinzy,chuchu\}@mit.edu}}%
}

}{
\pdfinfo{
   /Author (Homer Simpson)
   /Title  (Robots: Our new overlords)
   /CreationDate (D:20101201120000)
   /Subject (Robots)
   /Keywords (Robots;Overlords)
}
}

\begin{document}

\iftoggle{IROS}{
%%% IROS %%%
\thispagestyle{empty}
\pagestyle{empty}
}{
%%% RSS %%%
% You will get a Paper-ID when submitting a pdf file to the conference system
\author{Author Names Omitted for Anonymous Review. Paper-ID [add your ID here]}

%\author{\authorblockN{Michael Shell}
%\authorblockA{School of Electrical and\\Computer Engineering\\
%Georgia Institute of Technology\\
%Atlanta, Georgia 30332--0250\\
%Email: mshell@ece.gatech.edu}
%\and
%\authorblockA{Twentieth Century Fox\\
%\and
%\authorblockN{James Kirk\\ and Montgomery Scott}
%\authorblockA{Starfleet Academy\\
%San Francisco, California 96678-2391\\

% avoiding spaces at the end of the author lines is not a problem with
% conference papers because we don't use \thanks or \IEEEmembership

%Homer Simpson\authorrefmark{2},
%Montgomery Scott\authorrefmark{3} and
%Eldon Tyrell\authorrefmark{4}}
%Telephone: (800) 555--1212, Fax: (888) 555--1212}
%\authorblockA{\authorrefmark{4}Tyrell Inc., 123 Replicant Street, Los Angeles, California 90210--4321}}
}

\maketitle

\begin{abstract}
Reactive and safe agent modellings are important for nowadays traffic simulator designs and safe planning applications. In this work, we proposed a reactive agent model which can ensure safety without comprising the original purposes, by learning only high-level decisions from expert data and a low level decentralized controller guided by the jointly learned decentralized barrier certificates. Empirical results show that our learned road user simulation models can achieve a significant improvement in safety comparing to state-of-the-art imitation learning and pure control-based methods, while being similar to human agents by having smaller error to the expert data. Moreover, our learned reactive agents are shown to generalize better to unseen traffic conditions, and react better to other road users and therefore can help understand challenging planning problems pragmatically. 
% We give the guarantee for the system-wise safety and get strong empirical results on five different traffic datasets. In a case study, we also show our learned reactive agents can react better under challenging scenarios (e.g. crowded intersections).%At last, we also show how our reactive agents can be used for planning.
\end{abstract}

\iftoggle{IROS}{}{
%%% RSS %%%
\IEEEpeerreviewmaketitle
}

\section{Introduction}
% Safety is very important for traffic simulations.

% Traditional methods like RL learn from manually designed rewards , IRL learn from expert trajectories. They don't have guarantee for the safety.

% We proposed a novel way to plan for multi-agent system.

% Our reactive agent is also explainable.

% With this and that, we can give a guarantee that 

Understanding how different road participants (e.g. human drivers, pedestrians, cyclist) act in different traffic scenes plays a crucial role in developing modern self-driving techniques: It can give insights into how humans behave in each scenario, how to learn from human reactions, how to team autonomous and human agents, and can also provide a more realistic simulation environment for autonomous car developers. Most non-ego road participants in today's self-driving simulators (e.g. CARLA~\cite{dosovitskiy2017carla},  TORCS~\cite{wymann2000torcs}, SUMO~\cite{krajzewicz2010traffic}) can only ``replay" pre-defined trajectories or be guided by carefully designed cost functions and thresholds~\cite{best2017autonovi}. Such non-reactive or handcrafted agents are unrealistic, and therefore create a big gap between simulators and real-world scenarios.
%
% On the other side, machine learning techniques such as behavioral cloning~\cite{bain1995framework} and inverse reinforcement learning (IRL)~\cite{russell1998learning,ng2000algorithms} try to ``mimic" humans from the existing traffic data sets by cloning the possible actions of the agents at each state, or learning the underlying reward functions~\cite{bhattacharyya2018multi}. Those learning-based methods do not directly enforce some basic safety properties of the agents such as avoiding collisions, staying within lanes, and keeping speed below limits. Some IL-based \chuchucomment{IL is not introduced here. Maybe it is my lack of knowledge, but does IL and IRL assume each other?} methods encode undesirable traffic phenomena in the reward/cost functions~\cite{bhattacharyya2019simulating}. However, designing the proper reward/cost functions is tedious and non-trivial, and may lead to reward hacking~\cite{hadfield2017inverse}, %\chuchucomment{I feel the last sentence is weak but don't know a better one now}
%  where agents learn undesired behaviors to hack for high rewards~\cite{russell2002artificial,clark2016faulty}.
% \chuchu{
On the other side, imitation learning (IL) techniques~\cite{bain1995framework,russell1998learning,ng2000algorithms,bhattacharyya2019simulating,hadfield2017inverse} try to ``mimic" humans from the existing traffic datasets, by learning actions or reward functions instead of directly enforcing some basic safety properties of the road users (e.g. avoiding collisions, staying within lanes and  below speed limits). Therefore, the learned agents may have weird or unexplainable behaviors. See a more detailed discussion in the related works.
% }
% On the other side, imitation learning (IL) techniques such as behavioral cloning~\cite{bain1995framework} and inverse reinforcement learning (IRL)~\cite{russell1998learning,ng2000algorithms} try to ``mimic" humans from the existing traffic datasets. Those learning-based methods do not directly enforce some basic safety properties of the agents such as avoiding collisions, staying within lanes, and keeping speed below limits. Some IL-based  methods encode undesirable traffic phenomena in the reward functions~\cite{bhattacharyya2019simulating}. However, designing the proper reward functions is tedious and non-trivial, and may lead to reward hacking~\cite{hadfield2017inverse}, %\chuchucomment{I feel the last sentence is weak but don't know a better one now}
%  where agents learn undesired behaviors to hack for high rewards~\cite{russell2002artificial,clark2016faulty}.
%  \chuchucomment{IL is not introduced here. Maybe it is my lack of knowledge, but does IL and IRL assume each other?}\YM{According to GAIL: IL contains Behavioral Cloning and IRL, where the former is doing supervised learning and the latter learns a reward function that can reflect expert policy}

In this paper, we start from a different perspective and use a safety-driven learning-based control method to build reactive agent models for a variety of road users in different traffic scenarios. Control-theoretical tools exist to guide the synthesis of desired controllers using a certificate as the guidance. For example, control barrier functions (CBF) can be used to find controllers such that the closed-loop system satisfies safety properties defined by the corresponding barrier function~\cite{ames2014control,ames2019control}.
The biggest challenge of using CBF is the function design: Hand crafting a CBF for systems with complex dynamics and safety requirements is nearly impossible. Inspired by the recent advancement of learning decentralized CBF for the safety of homogeneous multi-agent systems~\cite{qin2021learning}, we want to learn only high-level decisions (e.g. when to switch the lane) from data and use neural CBF to guide the (low-level) safe actions of each agent for a group of heterogeneous road users.

Ideally, for a specific type of road user, a CBF that works across all scenarios would give a strong safety certificate. However, in practice, we notice that such a CBF is extremely difficult to learn, as the expected observations (or sensing inputs) of the agent vary significantly in different traffic scenarios. Therefore, we restrict our neural CBF to handle specific scenarios, where each of the traffic datasets can provide agents' observations in the corresponding scenario. This is analog to having different control policies when handling different road and traffic conditions.

% \chuchu{
We use the same learning framework to build simulation models for different road users in various traffic scenarios, including dense highway (NGSIM~\cite{simulation2007us}), normal highway (HighD~\cite{krajewski2018highd}), roundabouts (RounD~\cite{krajewski2020round}), mixed pedestrian-cyclist (SDD~\cite{robicquet2020learning}), and pedestrian-vehicle interaction at intersections (VCI-DUT~\cite{yang2019top})). We show that our safety-driven learning approach can significantly reduce the unsafe behaviors, comparing to both state-of-the-art IL-based methods, and traditional control methods. As a by-product, our learned road user simulations are similar to expert behaviors (i.e. those trajectories from the datasets when having the same starting positions and destinations) than IL-based methods, even when we do not directly optimize the similarity to the experts. Our learned simulation models can generalize well: Models trained using 20 agents can give similar safe results when used on up to 100 agents in the same scenario. Moreover, comparing to traditional model-based method like model predictive control (MPC), we achieve a 50X speedup when executing the model as we do not need to solve online optimization problems. In addition, we show that our agents react better in challenging traffic scenes: Our learned pedestrians will make ways (as normal pedestrians would do) to unfreeze the vehicle at crowded intersections.
The major contributions of our paper are:
\begin{inparaenum}
\item To our knowledge, we are the first to use decentralized neural CBF for heterogeneous multi-agent systems to build safe and reactive simulation models for road users from real-world traffic data.
\item Our CBF-based agents are proved to ensure safety, being even similar to the real-world behaviors (from the data) than leading IL-based methods, in a wide range of traffic scenarios.  
\item Our learned agents can generalize to unseen traffic conditions in the same scenario, and can react and coordinate with other road users in solving challenging planning problems.
\end{inparaenum}
All of above show the great potential of our proposed approach for building safe and realistic traffic simulations, which will benefit autonomous-driving planning applications. This is our first step of using safety as the primary goal when building road user simulation models. We seek to combine this idea with IL as the next step to further improve the similarity to actual humans. 
% }

% The major contributions of our papers are:
% \begin{inparaenum}
% \item We are the first to use decentralized CBF for heterogeneous multi-agent systems to build safe and reactive simulation models for various traffic scenarios.
% \item We conduct extensive experiments on five real traffic datasets (NGSIM~\cite{simulation2007us}, HighD~\cite{krajewski2018highd}, RounD~\cite{krajewski2020round}, SDD~\cite{robicquet2020learning} and VCI-DUT~\cite{yang2019top}) to demonstrate the superiority of our proposed approach over other strong baselines. \item We present how our reactive agents can coordinate with other user-controlled agent for challenging planning problems.
% All of above shows the great value of our proposed approach for future traffic simulations and for safe planning applications.
% \end{inparaenum}
\section{Related Work}

%Simulations platforms: CARLA, 

%Traffic datasets: Apollo, NuScene, HighD, NGSIM, RounD, NGSIM, SDD, VCI-DUT, 

% , there are several ways. First is rule-based, predefined , which is trivial and cannot reflect the dataset. Second is data-orientated, try to design some rules and . Recently, there is a trend to combine from both of the directions. And ours is aligned with the third direction.

%Rule-based agent modelling: There are some literature do rule-based modelling. ..., Like Model predictive Control, but it is very expensive. And from the control theory, can construct a predefined CBF to keep agents always in some safe set, hence ensure the safety. However, this needs lots of effort to define the cannot capture the real-world data, and also the policy is solved by some QP methods, which take quite a lot of time. 

\nitb{Imitation learning:} IL is a popular method for mimicking
% Much progress has been made in controlling  agents to mimic
expert's behaviors from demonstrations. 
Behavioral cloning~\cite{bain1995framework} learns the state-action pairs from the expert data via supervised learning, 
% which requires ample amount of data and does not generalize to unseen scenarios.
and inverse reinforcement learning~\cite{ng2000algorithms} seeks a cost function which prioritizes expert behaviors over other possible policies. Generative adversarial imitation learning (GAIL)~\cite{ho2016generative} uses a discriminator to distinguish expert's policy from non-expert behaviors generated by the policy network, and has been extended to share parameters~\cite{bhattacharyya2018multi} and augmented with rewards~\cite{bhattacharyya2019simulating} to discourage undesirable behaviors, for modeling multi-agent driving on highway.
% Recently, a parameter-sharing version of GAIL (PSGAIL)~\cite{} was proposed for multi-agent driver modeling on highway driving~\cite{simulation2007us}, and an augmented reward was used in RAIL~\cite{bhattacharyya2019simulating}. 
However, those methods do no directly enforce safety,
% don't have guarantees for the system's safety, 
which serves as a critical factor in traffic simulations. 
Model-free safe RL methods~\cite{liu2020robust,berkenkamp2017safe,cheng2019end,li2019temporal} can also be extended for multi-agent systems, but they draw less information from the expert trajectory. Moreover, finding the appropriate RL methods and reward functions are extremely challenging, and may lead to reward hacking, where agents learn undesired behaviors to hack for high rewards~\cite{clark2016faulty}.
% To take safety into account in RL, the first type of approach modifies the RL training paradigm by: 
% 1) reward-shaping~\cite{saunders2017trial}, 2) robust or Bayesian optimization~\cite{liu2020robust} and 3) add safety-related constraints for policy optimization~\cite{berkenkamp2017safe}. Though intuitive, they don't have any safety guarantees. The second type of approach combines RL with the control theory, using Lyapunov functions~\cite{chow2019lyapunov}, barrier certificates~\cite{cheng2019end,li2019temporal} to give safety guarantees. However, such methods often are computationally expensive and sensitive to controller structures and training algorithms. %And there are multi-agent collision avoidance reinforcement learning algorithms~\cite{}

% \chuchucomment{Move this here: However, designing the proper reward functions is tedious and non-trivial, and may lead to reward hacking, %\chuchucomment{I feel the last sentence is weak but don't know a better one now}
%  where agents learn undesired behaviors to hack for high rewards~\cite{clark2016faulty}}
 
\nitb{Safe control via barrier certificates:} Many learning methods are combined with barrier certificates~\cite{prajna2004safety} and CBF~\cite{wieland2007constructive, chen2017obstacle, ames2014control} to ensure safety.
ShieldNN~\cite{ferlez2020shieldnn} designs a NN controller guided by a given barrier function to rectify unsafe controllers' behaviors. \cite{wang2018permissive} uses Sum-of-Squares to learn a permissive barrier certificates and uses QP-based controller to ensure system's safety. \cite{srinivasan2020synthesis} uses SVM to model a CBF for safe control. \cite{long2020learning} constructs the CBF via objects' signed-distance function online to perform safe navigation. The frameworks for multi-agent CBF safe control are proposed in \cite{borrmann2015control} and \cite{wang2017safety}, and recently a decentralized multi-agent system using backup strategy is proposed by ~\cite{chen2020guaranteed}. The closest work to ours is the decentralized CBF for Multi-agent control ~\cite{qin2021learning}. But in ~\cite{qin2021learning}, all agents are of the same type and running the same simple reach-avoid controller, in an extremely simple environmental setup without real-world considerations.
% controllable under a simple simulation setting and no real world data is considered. 
Instead, our learned simulation model needs to handle real-world traffic scenarios and other users on the road, including other types of agents and agents that are non-reactive but only follow some pre-defined trajectories. Moreover, our controllers are also constructed in a hierarchical way for handling complex tasks.

\section{Preliminaries and Problem Statement}

% \subsection{Decentralized safe control and heterogeneous multi-agent system}
% Consider a heterogeneous multi-agent system: $\{x_1,x_2,...,x_N\}$ where the dynamics for the $i$-th agent follow a differential equation:
% \begin{equation}
%     \dot{x}_i(t)=f_i(x_i(t),u_i(t))
% \label{eq:system}
% \end{equation} 
% where $x_i(t)\in\mathcal{X}_i\subset\mathbb{R}^n$ denotes the agent's state information (pose, velocity, etc.) and $u_i(t)\in\mathcal{U}\subset \mathbb{R}^m$ denotes the actions or control inputs (acceleration, and angular velocity, etc.). 
%\chuchu{
We model the real-world traffic scenes as a heterogeneous multi-agent system in a specific scenario, where road participants (aka agents) choose their actions depending on their states and observations and follow some dynamics to move forward. A scenario defines what the road participants expect to observe.
%}
% Most traffic scenes in real world can be approximately modelled as a heterogeneous {\em multi-agent} system, where each road participant starts from an initial state and follows its corresponding dynamics. 
Formally, 
%\chuchu{
\begin{definition} \label{def:multi_agent_system}
A heterogeneous multi-agent system in a scenario is defined as a tuple $\mathcal{M} = \langle A_1,A_2,...,A_N, \scenario \rangle$, where 
\begin{enumerate}
    \item Each agent $A_i = \langle \mathcal{X}_i, \mathcal{U}_i, f_i, \mathcal{Z}_i \rangle$  is defined by its state space $\mathcal{X}_i \in \mathbb{R}^n$ (e.g. pose, velocity), input space $\mathcal{U}_i \subseteq \mathbb{R}^m$ (e.g. acceleration, and angular velocity), dynamic function $f_i: \mathcal{X}_i \times \mathcal{U}_i  \rightarrow \mathcal{X}_i$,  and observation space $\mathcal{Z}_i \in \mathbb{R}^{\ell_i}$. Moreover,
    \begin{itemize}
    \item The semantics of agent dynamics are defined by trajectories, which describe the evolution of states over time. Given an input trajectory $u_i: [0,\infty)\rightarrow \mathcal{U}_i$, then the state trajectory $x_i: [0,\infty)\rightarrow \mathcal{X}_i$ follows a differential equation:
        \begin{equation}
            \dot{x}_i(t)=f_i(x_i(t),u_i(t))
        \label{eq:system} 
        \end{equation} 
    \item At any time $t$, each agent $i$ can obtain a local observation $z_i(t) \in \mathcal{Z}_i \subset \mathbb{R}^{\ell_i}$, which contains sensing information of its surroundings such as close-by neighbors' position, lane markers, and static obstacles. 
    \item The agents are {\em reactive} as the control policies $u$ will be functions of the observations, which indicates that the agents react to the environmental changes.
    \end{itemize}
    \item A scenario $\scenario$ is defined as a set of possible values for the (state, observation) pairs of all agents: $\scenario = \{ \scenario_i \subseteq \mathcal{X}_i \times \mathcal{Z}_i \}_{i=1}^N$, such that for each agent $i$, the pair $(x_i, z_i)$ has to be contained in $\scenario_i$.
\end{enumerate}
% the dynamics for the $i$-th agent 
% here $x_i(t)\in\mathcal{X}_i\subset\mathbb{R}^n$ denotes the agent's state information (pose, velocity, etc.) and $u_i(t)\in\mathcal{U}_i\subset \mathbb{R}^m$ denotes the actions or control inputs (acceleration, and angular velocity, etc.). 
\end{definition}
%}
% \chuchu{
We introduce the notion of scenarios because the possible values for the  (state,observation) pairs vary significantly in diffenrent traffic scenes. 
For example, the possible observations of cars on the highway (which do not contain pedestrians or cyclists) are different from driving across the intersections (which do not contain high-speed cars). As we will see later in the experiment section, each dataset can be seen as a collection of samples from a specific scenario.
% }

Agents sharing the same dynamics $f_i$ and same structure of observations $z_i$ are called the same {\em type} of agents. For example, cars driving on the high-way can observe the leading and following cars in the same and neighboring lanes, and pedestrians at intersections can observe other pedestrians and cars that are not blocked by other agents. In this paper, we will develop a same control policy and safety certificate for the same type of agents. In what follows, we define the concept of safety for a multi-agent system in a given scenario, and how to enforce safety using decentralized barrier certificates.

The {\em safety} for the road participants means collision-free, staying with the lane, etc. As we are working on a heterogeneous system, the measurements for the safety are different among agents. Therefore, we use functions $\mu_i: \scenario_i \rightarrow \mathbb{R}$ to describe the safety. That is,  
% is defined
% The definition of {\em safety} for an agent is measured by a function of the agent's ego state and local observation: $\gamma_i:\mathcal{X}_i\times \mathcal{Z}_i \to\mathbb{R}$. 
the $i$-th agent is {\em safe} if $\mu_i(x_i,z_i)\leq 0$, otherwise the agent is {\em dangerous} (or {\em unsafe}).
% }

%\chuchucomment{Now you need to give an example of $\mu$s to give reader a better understand. You can use illustrative figures and simple scenarios to show that it means by x, z, and mu.}

The explicit form of $\mu_i$ could be varied for different types of agents: for example, the safety for pedestrians in crowds might be tested by the Euclidean distance measure from their locations (encoded in $x_i$) to their neighbors (encoded in $z_i$), and $\mu_i$ can be constructed as a predefined safety threshold minus this Euclidean distance. Whereas the safety checking for vehicles at intersections or roundabouts might require much more complicated functions, such as detecting the overlay of the vehicles' contours from the bird-eye-view. %The {\em safety} for the multi-agent system is defined by the conjunction of all agents' safety:
Fixing a control policy, the multi-agent system is {\em safe} if all of the agents in the system are {\em safe} in the given scenario:
\begin{equation}
    \mathcal{M} \text{ is safe} \Longleftrightarrow \max\limits_{i=1,..,N}(\mu_i(x_i,z_i))\leq 0 
    \label{eq:systemsafe}
\end{equation}

Control barrier functions (CBF)~\cite{ames2014control,ames2019control} are a control theoretical tool to find a safe controller and enforce the states of dynamic systems to stay within the safe set (in a forward invariant sense). Recently decentralized CBF~\cite{qin2021learning,chen2020guaranteed} are proposed to give decentralized controllers that can guarantee the safety of multi-agent systems. In this paper, we follow this idea and propose decentralized CBF for our multi-agent system in traffic scenes as defined in Definition~\ref{def:multi_agent_system}.
% }

% \chuchu{
\begin{proposition}\label{prop:cbf}
Given a multi-agent system in a scenario $\mathcal{M} = \langle A_1,A_2,...,A_N, \scenario \rangle$ as in  Definition~\ref{def:multi_agent_system}, if a function $h_i:\mathcal{S}_i \to \mathcal{R}$ which has the time derivative almost everywhere (denote as $a.e.\ t\in [0,\infty)$) and satisfies the following CBF conditions when the time derivative exists: 
\begin{equation}
\begin{cases}
   h_i(x_i,z_i) \geq 0,\,\forall (x_i,z_i)\in X_{i,s}  \\
   h_i(x_i,z_i) < 0,\,\forall (x_i,z_i)\in X_{i,d}  \\
   \dot{h}_i(x_i,z_i) + \alpha(h_i(x_i,z_i)) \geq 0 ,\, \forall (x_i,z_i)\in X_{i,+} ,
   \label{eq:cbf}
\end{cases}
\end{equation}
where  $ X_{i,s} = \{(x_i,z_i) \in \scenario_i \ | \ \mu_i(x_i, z_i) < 0 \}$ is the safe set for agent $i$,  $ X_{i,d} = \{(x_i,z_i) \in \scenario_i \ | \ \mu_i(x_i, z_i) \geq 0\}$ is the unsafe set for agent $i$, $X_{i,+}=\{(x_i,z_i)\ |\ h(x_i,z_i)\geq 0\}$, $\alpha$ is an extended class-$\mathcal{K}$ function which is defined as a strictly increasing continuous function $\alpha: (-b,a)\to(-\infty,\infty)$ for some $a,b>0$ and $\alpha(0)=0$, then we call $h_i$ a valid CBF for the scenario $\scenario$, and the safety of this multi-agent system is guaranteed as long as the initial state and observation pairs are in the safe set. 
% \chuchucomment{And now give the proofs.}
\end{proposition}

% \chuchucomment{You have to give the definition of extended class-K function somewhere as not every one knows about it.}
% }

% \begin{proof}
% We first prove the agent's safety guarantee, and then prove the guarantee for the system's safety. 

% 1. If the initial state $x_i(0)$ is in the safe set $X_{s,i}$, we prove $x_i(t)$ will never enter the dangerous set $X_{d,i}$, which guarantees the $i$-th agent's safety. 

% 1) When the function $h_i$ is differentiable everywhere, since $x_i(0)\in X_{s,i}$, we have $h_i(x_i(0),z_i(0))\geq 0$, $\forall z_i\in \mathcal{Z}_i$. Since $\dot{h}_i(x_i,z_i)+\alpha (h_i(x_i,z_i)) \geq 0$, the function $h_i$ will always be non-negative (the proof is provided in \cite{ames2014control}). This means $h_i \nless 0,\, \forall t>0$, therefore $(x_i(t),z_i(t))\notin \mathcal{X}_{d,i}\times \mathcal{Z}_i, \forall t>0$ and eventually we have $x_i(t)\notin X_{d,i},\forall t>0$

% 2) When the function $h_i$ is not differentiable everywhere, based on the Theorem 2 of \cite{glotfelter2017nonsmooth}, if the CBF conditions are satisfied under the generalized gradient, the function $h_i$ is still a valid CBF and guarantees the safety of the system.

% 2. Now if every agent in the system has the safety guarantee, from the definition in \eqref{eq:systemsafe}, we can show that this system's safety is guaranteed under scenario $\scenario$.
% \end{proof}

\begin{proof}
% \chuchu{
The proof for Proposition~\ref{prop:cbf} follows a standard procedure of proving safety by CBFs.
% }
We first prove the agent's safety guarantee, and then prove the guarantee for the system's safety.

1. If the initial state and observation pair $(x_i(0),z_i(0)) $is in the safe set $X_{i,s}$, we prove that $(x_i(t),z_i(t))$ will never enter the dangerous set $X_{i,d}$, thus ensures the agent's safety. 

%1) When the function $h_i$ is differentiable everywhere, since $(x_i(0),z_i(0))\in X_{i,s}$, we have $h_i(x_i(0),z_i(0))\geq 0$. Since $\dot{h}_i(x_i,z_i)+\alpha (h_i(x_i,z_i)) \geq 0$, the function $h_i$ will always be non-negative (the proof is provided in \cite{ames2014control}). This means $h_i \nless 0,\, \forall t>0$, therefore we have  $(x_i(t),z_i(t))\notin X_{i,d}, \forall t>0$.
Since $(x_i(0),z_i(0))\in X_{i,s}$, we have $h_i(x_i(0),z_i(0))\geq 0$. Let $\dot{y}_i(t)=-\alpha(y_i(t))$ and $y_i(0)=h_i(x_i(0),z_i(0))$. Because an extended class-$\mathcal{K}$ function $\alpha$ is local Lipschitz, solutions $y_i(t)$ exist and are unique, and according to~\cite[Lemma 4.4]{khalil2002nonlinear}, the solution is: $y_i(t)=\sigma(y_i(0),t)$ where $\sigma$ is a class $\mathcal{KL}$ function\footnote{A function $\sigma:\mathbb{R}_{\geq 0}\times \mathbb{R}_{\geq 0}\to\mathbb{R}_{\geq 0}$ is class-$\mathcal{KL}$ if it is class-$\mathcal{K}$ in its first argument and, for each fixed $r$, $\sigma(r,s)$ is strictly decreasing with respect to $s$ and $\lim\limits_{s\to\infty}\sigma(r,s)\to0$.}.  Now since $\dot{h}_i(x_i(t),z_i(t))\geq -\alpha(h_i(x_i(t),z_i(t))),\, a.e.\, t\in [0,\infty)$, we have $h_i(x_i(t),z_i(t))\geq y_i(t),\, \forall t\in[0,\infty)$ by~\cite[Theorem 1.10.2]{lakshmikantham1969differential}. Thus $h_i(x_i(t),z_i(t))\geq \sigma(y_i(0), t)=\sigma(h_i(0),t)\geq 0,\forall t\in [0,\infty)$. So we have $(x_i(t),z_i(t))\notin X_{i,d}, \forall t>0$. 

%2) When $h_i$ is not differentiable everywhere, we can define $\dot{h}_i$ as the generalized gradient introduced in~\cite{glotfelter2017nonsmooth}, and based on~\cite[Theorem 3]{glotfelter2017nonsmooth}, if the CBF conditions are satisfied under the generalized gradient, the function $h_i$ is still a valid CBF, hence guarantees the safety of the system.

2. Now if every agent in the system has the safety guarantee, from the definition in \eqref{eq:systemsafe}, we can show that the safety of this system is guaranteed under scenario $\scenario$.
\end{proof}

% \chuchu{
Notice that the decentralized CBF in Proposition~\ref{prop:cbf} is only defined on the given scenario $\scenario$. This is because handcrafting a CBF for a complex heterogeneous multi-agent system is impossible, while learning a single valid CBF for all possible scenarios is extremely difficult. The latter requires the CBF to be powerful and expressive enough to handle any traffic condition and therefore hard (if not impossible) to learn. In this paper, we take a middle-ground and only require the agents to learn CBFs that are valid for each scenario and therefore guarantee the safety only for the specific scenario.
% }

\section{Build safe and reactive road participants models through learning neural CBF.}
\begin{figure}[htb]
\includegraphics[width=0.5\textwidth]{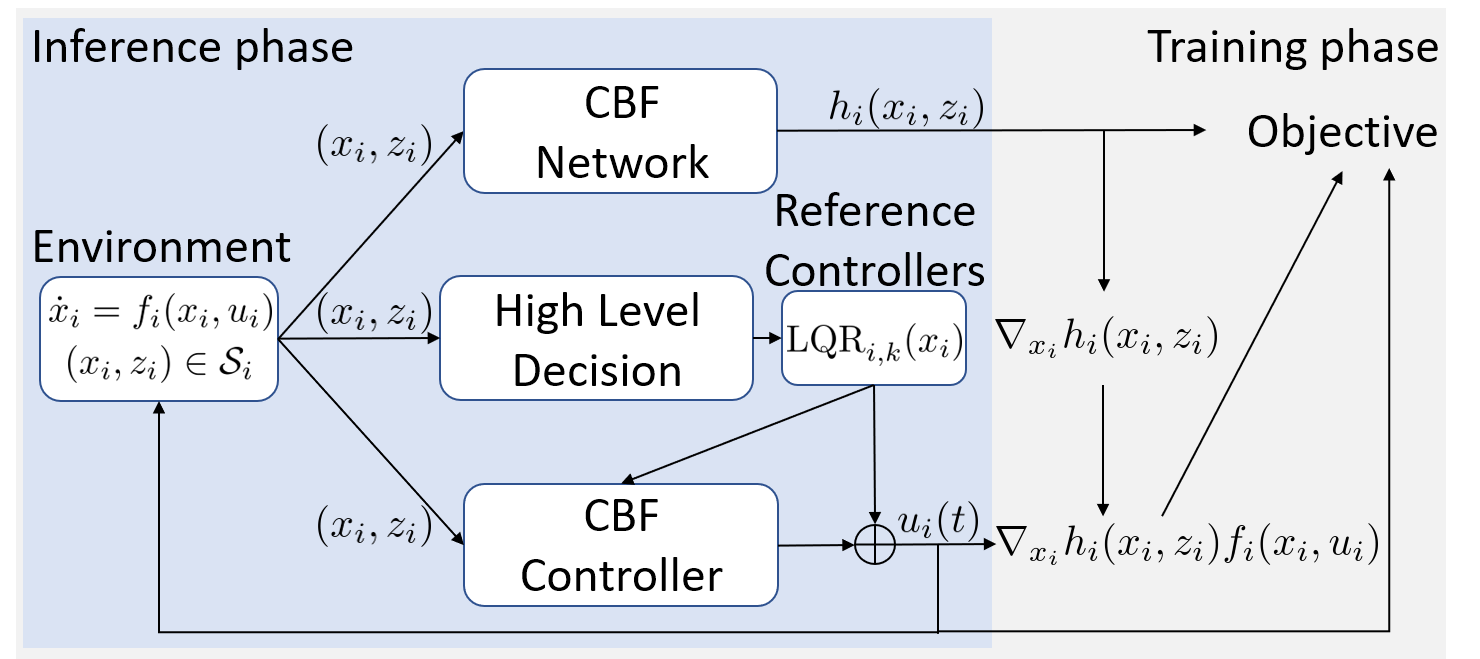}
\centering
\caption{Overview of the learning framework for reactive agents. Under each scenario, each type of agents share the same learning framework and control policies. %The $i$-th agent's model consists of a learned CBF network, a learned high-level decision, reference controllers and a learned CBF controller. 
}
\label{fig:diagram}
\end{figure}

Given the multi-agent system as in Definition~\ref{def:multi_agent_system} and guided by the Proposition~\ref{prop:cbf}, we propose a decentralized learnable framework to construct autonomous reactive agents with safety as the primary objective. As shown in \figref{fig:diagram}, for each type of agents under the shared simulation environment, we use a hierarchical control framework that contains a high-level decision/controller, a set of reference controllers, a CBF network (served as $h_i$ in~\eqref{eq:cbf}) and a CBF-guided controller. The high-level controller is pre-trained offline using expert trajectories to capture high-level decisions made by humans that are hard to defined in CBF (e.g. when to switch lanes). Then during training in the simulation, the CBF network and CBF controller jointly learn to satisfy the CBF conditions in~\eqref{eq:cbf}. 
At the inference stage, the high-level controller selects a reference controller for different purposes (e.g. staying at the center of the lane, reaching destination, etc.), and the CBF controller rectifies the reference control command to ensure system's {\em safety} while avoids diverging too much from the agents original purposes, hence being realistic. 
% At the inference stage, the high-level controller selects a reference controller catering to environment and agent's observation, and the CBF controller rectifies the reference control command to ensure system's {\em safety} while avoids diverging too much from \chuchu{the reference controller}, hence being realistic. 
Detailed implementations and training procedures will be presented in the rest of this section.

%Given the state observation pair from the environment, the high-level decision selects a specific reference controller, then the CBF controller further rectifies it to be the final control for the agent. In training, the CBF Network and CBF controller are jointly learned to satisfy the CBF conditions defined in \eqref{eq:loss} to ensure that at the inference stage, the learned CBF controller can maintain the system safety.
\subsection{Reference controllers and high-level controllers}
% To execute more realistic and meaningful maneuvers like driving in a roundabout, we use reference controllers synthesized by classical approaches. 
CBF only enforces safety and therefore could lead to overly safe actions. For example, agents using only CBF-based control could choose driving off the road to stay far from others. Therefore, we need a reference controller to serve the original purposes of the agents.
We adopt linear quadratic regulator (LQR) as the reference control in our experiments, but it can be any other forms. 
Moreover, some high-level decisions are intentions of humans and thus better to be learned from data. For example, whether to switch a lane on the highway given the current traffic condition. Therefore, we further construct a high-level controller for scenarios that would need such a high-level  decision for intentions.
% contain different patterns of behaviors, like lane-keeping and lane-changing. 
Here the high-level controller $Q_i:\mathcal{X}_i\times \mathcal{Z}_i\to \mathbb{N}_i$ returns a discrete value that represents the behavior pattern or decision, based on which the corresponding reference controller $\text{LQR}_{i,k}: \mathcal{X}_i\to\mathcal{U}_i$ is chosen.
%to select from different reference controllers. 
Finally, the CBF controller $\pi_{i,0}(x_i,z_i)$ learns to rectify the reference control commands to maintain system's safety. The overall control input for the agent thus can be written as:
\begin{equation}
    u_i(x_i, z_i) = \sum\limits_{k\in \mathbb{N}_i}\mathds{1}(Q_i(x_i,z_i)=k)\cdot\text{LQR}_{i,k}(x_i) + \pi_{i,0}(x_i,z_i)
    \label{eq:control}
\end{equation}

\subsection{Agent safety checking decomposition and vectorized CBF}\label{subsec:vec}
Ideally, if a learnable barrier certificate together with a learnable controller can satisfy the conditions in \eqref{eq:cbf} for a system $\mathcal{M}$ , a scenario $\scenario$ and sets $X_{i,s},X_{i,d}$, then the safety of $\mathcal{M}$ can be fulfilled. However, in application the observation $z_i$ often lies in a high dimension space, leading learning barrier certificates from data a challenging problem. 

Notice that checking the agent safety in the system normally can be further decomposed to pairwise checkings between the agent and each of its neighbors or static obstacles~\cite{chen2020guaranteed}. For example, in a crowd scenario, a pedestrian is safe if it doesn't collide with any of its neighbors or static facilities. The pairwise checking with neighbors will render more information (labels) than just checking the agent safety in the learning process. Inspired by this, we propose a method to efficiently solve this by using a vectorized version of CBF, which (if learned correctly) can still guarantee system's safety.

For each agent, we assume the observation $z_i$ can be decomposed to the instance-specific observations with respect to its neighboring agents and other static obstacles, i.e. $z_i=(\tilde{z}_{i,1}, \cdots, \tilde{z}_{i,\mathcal{N}_{i}})$, where $\tilde{z}_{i,k}\in\tilde{\mathcal{Z}}_{i,k}$ and $\mathcal{N}_{i}$ denotes the maximum number of instances the $i$-th agent can perceive from its local observation at one time. The safety for an agent-neighbor pair is measured by the agent's state and its instance-specific observation: $\tilde{\mu}_{i,k}:\mathcal{X}_i\times \tilde{\mathcal{Z}}_{i,k}\to\mathbb{R}$. The $i$-th agent with its $k$-th neighbor is in {\em safe} set $\tilde{X}_{i,k,s}$ if  $\tilde{\mu}_{i,k}(x_i,\tilde{z}_{i,k})\leq 0$, otherwise is in {\em dangerous} set $\tilde{X}_{i,k,d}$

We denote function $H_i:\mathcal{X}_i\times \mathcal{Z}_i \to \mathbb{R}^{\mathcal{N}_i}$ in the form of $H_i(x_i, z_i)=[\tilde{H}_{i,1}(x_i, \tilde{z}_{i,1}),\cdots, \tilde{H}_{i,\mathcal{N}_i}(x_i, \tilde{z}_{i,\mathcal{N}_i})]^T$. Now if $\forall k$, $\tilde{H}_{i,k}$ satisfies the CBF condition defined in \eqref{eq:cbf} under corresponding sets $\tilde{X}_{i,k,s}$, $\tilde{X}_{i,k,d}$ and $\tilde{X}_{i,k,+}=\{(x_i,\tilde{z}_{i,k})|\tilde{H}_{i,k}(x_i,\tilde{z}_{i,k})\geq 0\}$, %\ZY{$\tilde{h}_{i,|\mathcal{N}_i|}(x_i, z_{i,|\mathcal{N}_i|})]^T$, should we use $N_i$ or $|\mathcal{N}_i|$ to represent the number of neighbouring agents? I think $N_i$ is less redundant than $|\mathcal{N}_i|$}
then we can construct  $h_i=\min\limits_{k}(\tilde{H}_{i,k})$, which will satisfy the CBF conditions defined in \eqref{eq:cbf} on $X_{i,s}$ and $X_{i,d}$. This can be proved using the Proposition 5 from~\cite{glotfelter2017nonsmooth} which discusses the sufficient conditions to construct a valid barrier function via min/max operations on top of component barrier functions.

Thus we can decompose agent's observation to instance-specific observations, and learn a collections of ``smaller" CBF to guarantee the system safety. In this way, the dimension for the input space for each CBF is greatly reduced and therefore easier to be learned. In addition, for the same type of neighbor agents, this method will provide $(K_{i,j}-1)$ more training samples to the corresponding CBF network, where $K_{i,j}$ is the average number of $j$-th type of neighbor agents that can be observed by the $i$-th agent in the scenario, hence greatly accelerates the learning process.

\subsection{Neural network implementations %\ZY{Why does the NN network section explain the theory? Should it explain the NN architecture? And I'm confused with the vectorized CBF defined in the following paragraph.}\YM{Moved to the previous section}
}

In this paper, the high-level controller, the barrier certificate $h_i$ and the CBF controller $\pi_{i,0}$ are constructed as neural networks (NN). The high-level policy is a Gated Recurrent Network which takes the agent's state and observation as inputs, and outputs discrete high-level decisions. As for the CBF controller, inspired by~\cite{qin2021learning}, we use a PointNet~\cite{qi2017pointnet} structured network to ensure the NN is permutation and length invariant to inputs, and outputs the control signal $\pi_{i,0}(x_i(t),z_i(t))$. %\ZY{PointNet is not used in Yue's implementation. We may remove that.}\YM{The figure above may be misleading - I still used a max-pooling of features in the middle of the policy network, and then concat to several fully-connected layers to output the control}
The CBF network is a multilayer perceptron, which takes the agent's state and observation $(x_i, z_i)$ as input and outputs a vector representing the neighbor-wise CBF value: $H_i(x_i, z_i) \in \mathbb{R}^{\mathcal{N}_i}$ as discussed in Section~\ref{subsec:vec}.

\begin{figure*}[htb]
\begin{subfigure}[b]{0.19\textwidth}
\includegraphics[width=0.99\textwidth]{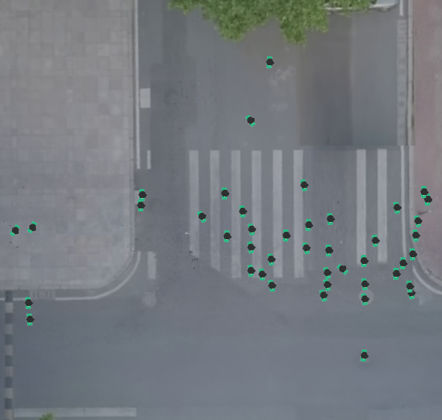}
\caption{\footnotesize Screenshot of VCI-DUT.}
\end{subfigure}
\begin{subfigure}[b]{0.22\textwidth}
\includegraphics[width=0.99\textwidth]{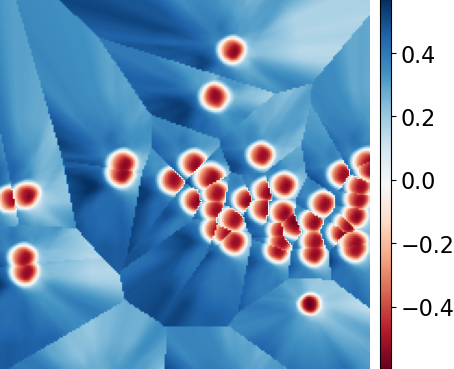}
\caption{\footnotesize CBF heatmap on VCI-DUT.}
\end{subfigure}
\begin{subfigure}[b]{0.28\textwidth}
\includegraphics[width=0.99\textwidth]{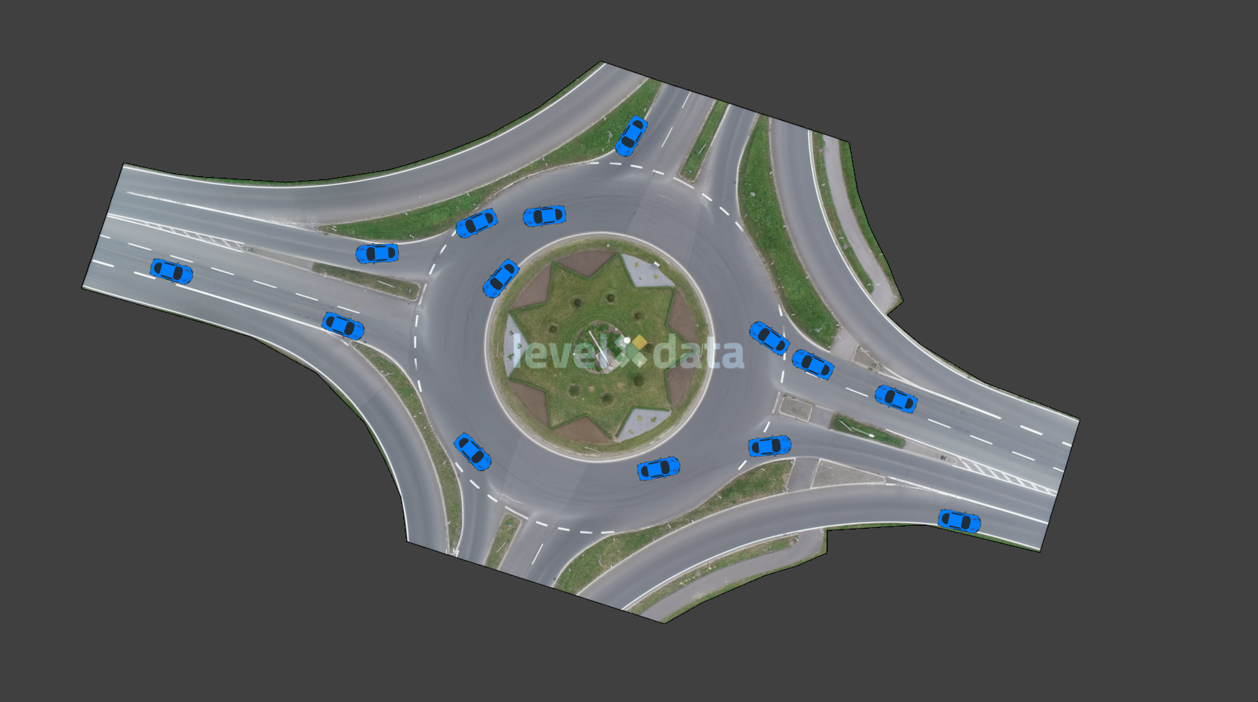}
\caption{\footnotesize Screenshot of RounD.}
\end{subfigure}
\begin{subfigure}[b]{0.29\textwidth}
\includegraphics[width=0.99\textwidth]{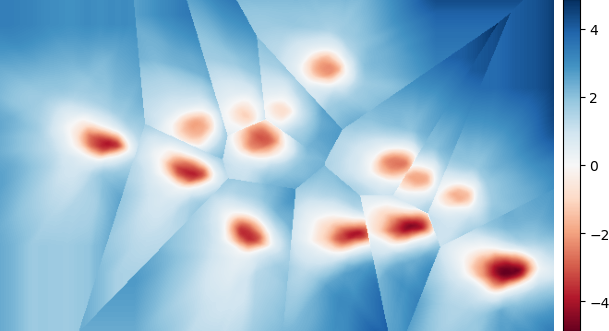}
\caption{\footnotesize CBF heatmap on RounD.}
\end{subfigure}

\centering
\caption{CBF visualization on VCI-DUT/RounD datasets.}
\label{fig:cbf-vis}
\end{figure*}

%\ZY{We may need some explanation of the relationship between the 3 terms in the loss function and the 3 CBF conditions.}\YM{did some interpretations below}
% The overall loss function for training $h_i$ and $\pi_i$ can be represented as:
We first train the high-level controller from a collection of expert trajectories with manually labelled high-level decisions. During the training phase in simulation, agents controlled by~\eqref{eq:control} will sample trajectory data containing agent's state, observation and action from each time step. To optimize the CBF networks $H_i$ and the controllers $\pi_{i,0}$ for satisfying the CBF conditions defined in \eqref{eq:cbf} on those sampled trajectories, we introduce the following loss function:
% \begin{equation}
% \begin{aligned}
% \mathcal{L} & = \sum\limits_{(x_i,z_i)\in X_{i,s}}\mathcal{L}_1(x_i,z_i) + \alpha \sum\limits_{(x_i,z_i)\in X_{i,d}} \mathcal{L}_2(x_i,z_i) \\
% & + \beta \sum\limits_{(x_i,z_i)\in X_{\geq 0}}\mathcal{L}_3(x_i,z_i)
% \end{aligned}
% \end{equation}

\begin{equation}
\begin{aligned}
\mathcal{L}_{\text{CBF}} & = \mathcal{L}_1 + \mathcal{L}_2 + \mathcal{L}_3
\end{aligned}
\label{eq:loss}
\end{equation}
% with
% \begin{equation}
% \begin{cases}
%     \mathcal{L}_1(x,z) =& \text{ReLU}(\gamma - h_i(x_i,z_i; \theta_i) \\
%     \mathcal{L}_2(x,z) =& \text{ReLU}(0,\gamma+h_i(x_i, z_i;\theta_i) \\
%     \mathcal{L}_3(x,z) =& \text{ReLU}(\gamma-\nabla_{x_i}h_i\cdot f_i(x_i,\pi(x_i,z_i;\omega_i))\\
%     &-\nabla_{z_i}h\cdot \dot{z}_i-\alpha(h_i)) \\
% \end{cases}
% \end{equation}
where:
% \begin{equation}
% \begin{cases}
%     \mathcal{L}_1 =& \sum\limits_{(x_i,z_i) \in X_{i,s}}\text{ReLU}(\eta - h_i(x_i,z_i)) \\
%     \mathcal{L}_2 =& \sum\limits_{(x_i,z_i) \in X_{i,d}}\text{ReLU}(\eta+h_i(x_i, z_i)) \\
%     \mathcal{L}_3 =& \sum\limits_{(x_i,z_i) \in \scenario}\text{ReLU}(\eta-\nabla_{x_i}h_i\cdot f_i(x_i,\pi(x_i,z_i))\\
%     &-\nabla_{z_i}h_i\cdot \dot{z}_i-\alpha(h_i)) \\
%     %\mathcal{L}_4 =& \sum\limits_{(x_i,z_i)\in X_{i,s}} ||\pi_i(x_i,z_i)||_2^2
% \end{cases}
% \end{equation}
\begin{equation}
\begin{cases}
    \mathcal{L}_1 =& \sum\limits_{(x_i,\tilde{z}_{i,k}) \in X_{i,k,s}}\text{ReLU}(\eta - \tilde{H}_{i,k}(x_i,\tilde{z}_{i,k})) \\
    \mathcal{L}_2 =& \sum\limits_{(x_i,\tilde{z}_{i,k}) \in X_{i,k,d}}\text{ReLU}(\eta+\tilde{H}_{i,k}(x_i, \tilde{z}_{i,k})) \\
    \mathcal{L}_3 =& \sum\limits_{(x_i,z_i) \in \scenario_i}\text{ReLU}(\eta-\nabla_{x_i}\tilde{H}_{i,k}\cdot f_i(x_i,\pi(x_i,z_i))\\
    &-\nabla_{\tilde{z}_{i,k}}\tilde{H}_{i,k}\cdot \dot{\tilde{z}}_{i,k}-\alpha(\tilde{H}_{i,k})) \\
    %\mathcal{L}_4 =& \sum\limits_{(x_i,z_i)\in X_{i,s}} ||\pi_i(x_i,z_i)||_2^2
\end{cases}
\end{equation}
here $\text{ReLU}(x)$ stands for the function $\max(x,0)$, and $\eta$ is a margin to enforce training stability. %The first three terms in \eqref{eq:loss} reflect the CBF conditions defined in \eqref{eq:cbf}: 
 $\mathcal{L}_1$ and $\mathcal{L}_2$ guide the neural barrier certificates $\tilde{H}_{i,k}$ to correctly distinguish samples $(x_i,\tilde{z}_{i,k})$ from the safe and dangerous sets $X_{i,k,s}, X_{i,k,d}$ respectively, and  $\mathcal{L}_3$ leads the controller and neural barrier certificates to satisfy the derivative condition $\dot{\tilde{H}}_{i,k}+\alpha (\tilde{H}_{i,k})\geq 0$.  Due to the difficulties of directly computing the $\nabla_{\tilde{z}_{i,k}} \tilde{H}_{i,k}\cdot \dot{\tilde{z}}_{i,k}$ term in $\dot{\tilde{H}}_{i,k}$, we approximate $\dot{\tilde{H}}_{i,k}$ by numerical differentiation\footnote{In this paper, we assume observations are noise-free; the numerical derivative approximation might be inaccurate for applications with noisy sensor data and we leave it for future works.}: $\dot{\tilde{H}}_{i,k}(x_i(t), \tilde{z}_{i,k}(t))\approx (\tilde{H}_{i,k}(x_i(t+\Delta t), \tilde{z}_{i,k}(t+\Delta t))- \tilde{H}_{i,k}(x_i(t), \tilde{z}_{i,k}(t)))/\Delta t$, where $\Delta t$ is the time interval for each simulation step. To sum up, $\mathcal{L}_{\text{CBF}}$ guides the control barrier certificates and the controller to ensure system safety.

%\footnote{We assume observations are noise-free; this approximation is inaccurate for applications with noisy sensor data and we leave it for future works.}

To avoid the CBF controller deviating too far from a normal behavior characterized by the reference controllers, we define a regularization term: $\mathcal{L}_r = \sum\limits_{(x_i,z_i)\in X_{i,s}} ||\pi_{i,0}(x_i,z_i)||_2^2$ 
%\ZY{in eqn 4 we used $\pi_{i,0}(x_i,z_i)$, but now it is $\pi_{i}(x_i,z_i)?$}
to penalize for the large output values of the controller when the sample $(x_i,z_i)$ is in the safe set $X_{i,s}$. Thus the total objective function becomes:
$
    \mathcal{L}= \mathcal{L}_{\text{CBF}} + \beta \cdot \mathcal{L}_r
$,
where $\beta$ balances the weights between system's safety and other agent purposes, and is set to 0.1 in our experiments.

% \subsection{Visualization for the Control Barrier Functions}
% To illustrate what has been learned by our CBF network, we pick scenarios from the VCI and RounD datasets and plot the CBF value for each location in the scene as if a new agent were assigned to that spot. As shown in \figref{fig:cbf-vis}, under both scenarios our CBF network outputs negative value when the location is close to other road participants (which is dangerous), and shows positive value at places far away from other agents (safe). In the RounD case, even if the neural network does not learn a perfect CBF that exactly reflects contours of neighboring vehicles, as later on shown in \figref{fig:four}, our method can still achieve the lowest collision rate\footnote{Details about the VCI and RounD datasets and the collision rate will be presented in Section \ref{sec:experiments}.}, which shows the robustness of our CBF learning framework.
To illustrate what can be learned using the above framework, we demonstrate contour plots of the learned CBF value in two scenarios (VCI-DUT and RounD datasets, to be discussed in detail in Section~\ref{sec:experiments}) for pedestrians and cars respectively. 
As shown in \figref{fig:cbf-vis}, the contour plots show the value of a new agent's CBF for each location as if that agent were assigned to be in the scene.
We see that under both scenarios, our CBF network outputs negative value when the location is close to other road participants (which is dangerous), and shows positive value at places far away from other agents (safe). Therefore our learned agents will prefer to go to locations that are ``bluer".

\section{
% \chuchu{
Simulation models for real-world scenarios
% }
}
\label{sec:experiments}
\begin{figure*}[htb]
\begin{subfigure}[b]{0.195\textwidth}
\centering
\includegraphics[width=0.98\textwidth]{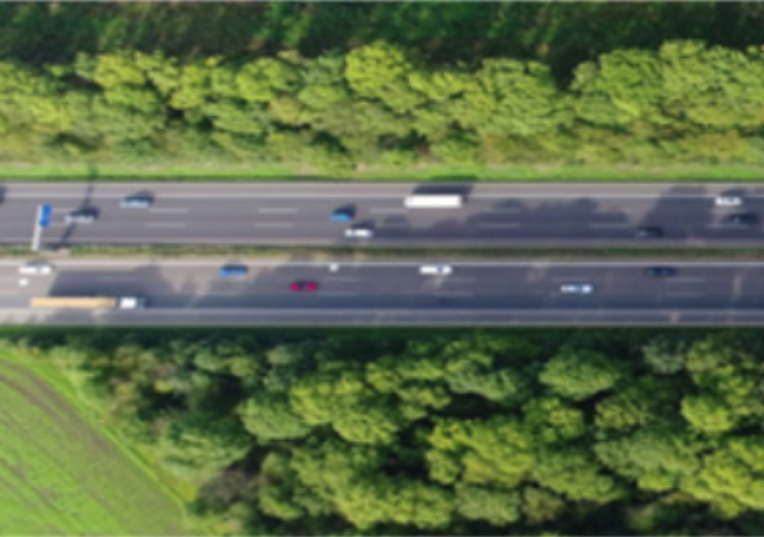}
\caption{\footnotesize {\highd} contains $>$110500 vehicle trajectories at six different locations on German highways.}
\end{subfigure}
%\\\\
\begin{subfigure}[b]{0.195\textwidth}
\centering
\includegraphics[width=0.98\textwidth]{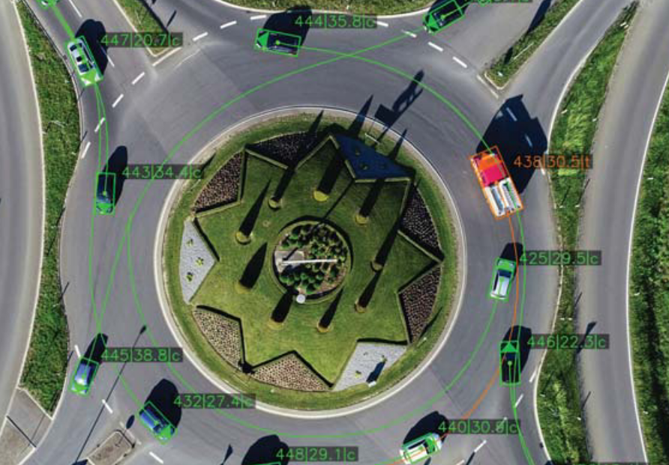}
\caption{\footnotesize {\round} contains $>$6 hours of trajectories from 13746 road users at $3$ different German roundabouts.}
\end{subfigure}
%\\\\
\begin{subfigure}[b]{0.195\textwidth}
\centering
\includegraphics[width=0.98\textwidth]{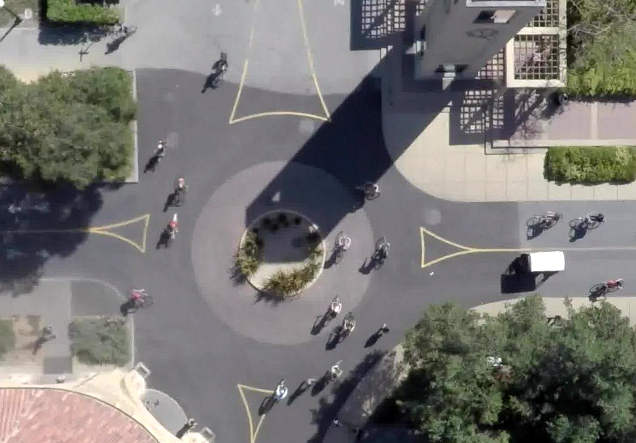}
\caption{\footnotesize {\sdd} records pedestrians and bicyclists trajectories at $8$ recording sites on Stanford campus.}
\end{subfigure}
%\\\\
% \includegraphics[width=0.15\textwidth]{figs/demoexp/00052.png} \hfill
\begin{subfigure}[b]{0.195\textwidth}
\centering
\includegraphics[width=0.98\textwidth]{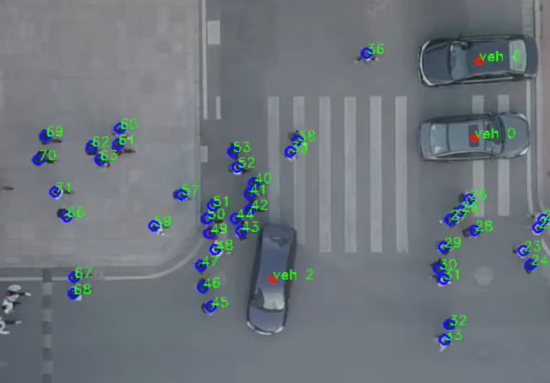}
\caption{\footnotesize {\vci} contains 1793 pedestrians trajectories interacting with vehicles at two intersections.}
\end{subfigure}
%\\\\
\begin{subfigure}[b]{0.195\textwidth}
\centering
\includegraphics[width=0.98\textwidth]{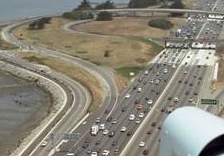}
\caption{\footnotesize \ngsim contains 45 minutes of highway driving data on US Highway 101 and Interstate 80.}
\end{subfigure}
\centering
\caption{Real world traffic datasets used in this paper.}
\label{fig:screenshots}
\end{figure*}

We demonstrate the performance of our approach in various simulations as shown in \figref{fig:screenshots}.
% We \chuchu{demonstrate in simulation of the performance of our approach.} 
% conduct extensive experiments to show that our algorithm can provide strong safety guarantee in challenging simulations with real world traffic data.
% \chuchu{
% We used five traffic datasets that contain a mixture of difference road users and under different scenarios. The details of the datasets can be found in \figref{fig:screenshots} and the corresponding references. 
% }
% \chuchucomment{Summarize comparison results using 1 or 2 sentences.}
Comparing to other baselines, our method achieves the lowest collision rate and RMSE, being 50X faster in terms of execution time than model-based approach like MPC, and can generalize and react reasonably to unseen conditions to help solve challenging planning problems. 
% Our reactive agent models can also help for user-controlled agents with planning under challenging scenarios.

% \subsection{Datasets}

%, namely the Highway Drone Dataset (HighD), the Roundabouts Drone Dataset (roundD), the Stanford Drone Dataset (SDD), the Vehicle-Crowd Interaction Dataset (VCI-DUT) and the Next-Generation Simulation (NGSIM) dataset.

% We used five traffic datasets (shown in \figref{fig:screenshots}) as followed: 

% \nitb{\highd:} The HighD dataset contains more than 110500 vehicle trajectories recorded at 25Hz at six different locations on German highways.

% \nitb{\round:} The RounD dataset records more than 6 hours of trajectories from 13746 road users (mostly cars and trucks) at three different roundabouts in German.

% \nitb{\sdd:} The Stanford drone dataset records road user trajectories at eight recording sites on campus, where the majority are pedestrians and bicyclists.

% \nitb{\vci:} The VCI-DUT dataset was collected on campus where the road participants (pedestrians, vehicles) can move freely without any specific priority.

% \nitb{\ngsim:} The Next-Generation Simulation dataset contains 45 minutes of highway driving data recorded at 10Hz for US Highway 101 and Interstate 80.

\subsection{Implementation details, baseline and metrics}

\nitb{Agent types and dynamics:} We categorize all agents into three major types: vehicles, cyclists and pedestrians. For vehicle-type and cyclist-type agents, we adopt a unicycle model with different constraints, whereas a double integrator model is used for pedestrian-type agents. An agent can only obtain observations within 30 meters of its surroundings.

\nitb{Safety measures:} The pair-wise safety measure function between vehicles is defined as the area of intersection between the vehicles' contours from the bird-eye-view. The pair-wise safety measure function among all other combinations of agent types is defined as: $r_1+r_2-d$ where $r_1, r_2$ are the radius for the minimum cover circle of the agents' contours, and $d$ is the Euclidean distance between two agents.

\nitb{Reference / high-level controllers:} 
LQR is used to give the reference control policy for the following purposes:
\begin{inparaenum}
\item For SDD and VCI-DUT (pedestrian, cyclists and vehicles in free space/campus), LQR is designed for reaching the destination.
\item For RounD (vehicles at roundabouts), LQR is used on vehicles to keep tracking on the pre-defined road points to enter/exit and drive in the roundabout. 
\item For HighD and NGSIM (vehicles on highways), LQR is designed for lane-keeping or lane-switching to the adjacent lane. 
\end{inparaenum}
High-level controllers are only used on HighD and NGSIM, where we learn from data when to switch the lane. This is because there could exist a safe controller no matter whether the vehicle switches the lane or keeps in the current lane. The lane-switch decision learned from data can better capture the intention of human drivers. For other datasets, the pedestrians and cyclists just move towards their pre-defined destinations.

% We adopt different LQRs as reference controllers for different scenarios. For SDD and VCI-DUT datasets (pedestrian, cyclists and vehicles in free space/campus), the LQR is designed for goal reaching. For RounD (pedestrian, cyclists and vehicles at roundabouts), we design LQR for vehicles to keep tracking on the predefined road points to enter/exit the roundabout or to drive in the roundabout. For HighD and NGSIM (vehicles on highways), the LQR is designed for keeping on the centerline of the current lane or switching to the centerline of the adjacent lane. We only tried high-level controllers on HighD and NGSIM where it is plausible to label different modes of the driving behaviors (lane-keeping and lane-changing). 
% %For those two datasets, we design the LQR to keep tracking on the centerline of the lane. 
% If a lane-changing behavior is triggered by the high-level controller, we change the LQR to track on the adjacent lane, until the lane-changing maneuver is succeeded. Otherwise, the LQR will  let vehicles drive in the current lane.

\nitb{Training and testing:} In each simulation run, we control 10$\sim$100 agents in the scene for 5$\sim$20 seconds (equivalent to 50 to 200 time steps), and let the rest agents in the scenario follow the expert trajectories. Training on a subset of one scenario for our approach takes 1$\sim$3 hours on an RTX 2080 Ti graphics card. During testing, we start from a different split of the dataset in the same scenario, perform 100 testing runs and average the metrics discussed below. For NGSIM, we follow the same testing configurations reported in \cite{bhattacharyya2018multi}.

% For HighD and NGSIM, we use a high-level policy network to make lane-keeping and lane-changing decisions, and then use the corresponding reference controller to execute the maneuver. For RounD, we pre-define a primal trajectory for vehicles to enter/exit the roundabout. For SDD, two different CBF and controllers are learned for pedestrians and cyclist.  For both SDD and VCI, the reference controller will instruct the agents to reach the destinations. In all of different scenarios. the CBF controllers are learned to guarantee the agents' safety. For those model-based methods, the reference controllers are provided.

% \paragraph{Baselines}
% For HighD, RounD, SDD and VCI-DUT, we compared our algorithm with Behavior Cloning~\cite{bain1995framework}, PS-GAIL~\cite{bhattacharyya2018multi}, MPC~\cite{garcia1989model} and LQR.
\noindent \textit{\underline{Baseline approaches.}}
For HighD, RounD, SDD and VCI-DUT datasets, we compare our approach with behavioral cloning (BC)~\cite{bain1995framework}, PS-GAIL~\cite{bhattacharyya2018multi} and MPC~\cite{garcia1989model}.
On the NGSIM dataset, we compare with the results of PS-GAIL~\cite{bhattacharyya2018multi} and RAIL~\cite{bhattacharyya2019simulating} reported in their paper. We do not report the results of RAIL~\cite{bhattacharyya2019simulating} on the other datasets as it is too sensitive to the handcrafted rewards. Details of the implementation of these baseline methods are as follows:
\textbf{Behavioral cloning} is trained on the ``state-observation, action" pairs from the expert trajectories using L2 loss (comparing to the ground-truth control signals) until converged, using a fully-connected NN. 
% We used the same input, output as the CBF controller and a fully-connected neural network for the best performance.
\textbf{PS-GAIL} results on NGSIM are reproduced from the original implementation as reported in~\cite{bhattacharyya2018multi}. For the rest datasets, we follow the network structure reported in~\cite{bhattacharyya2018multi} and train in simulations for the same number of epochs as for our method.
\textbf{MPC} is used to replace the neural-CBF controllers for safety, and the same pre-trained high-level controller and reference controllers are used. We use Casadi~\cite{Andersson2019} for MPC solving, with a goal-reaching cost and collision-avoidance constraints. The constraints are relaxed by slack variables to ensure the feasibility of the problem. The MPC results reported are fine-tuned results. 
\textbf{RAIL} employs a handcrafted augmented reward in imitation learning, and provides the best results on collision rate, off-road duration and hard brake rate on the NGSIM dataset in published literature. We use the original implementation of RAIL provided by the authors.
Lastly, we compare with a state-of-the-art (none-learning) calibration methods \textbf{$\text{IDM}_\theta$}~\cite{bhattacharyya2020online} on NGSIM and HighD. Again the $\text{IDM}_\theta$ models in~\cite{bhattacharyya2020online} could only work for highway car-following cases so we do not compare with it on other datasets.
% \nitb{LQR:} We used the same pretrained high-level controller and reference controllers as in our framework and removed the CBF controller. Hence no training in simulation is needed for this baseline.

% \nitb{MPC:} Based on the LQR baseline, we solved a finite time optimal control problem for each agent with a goal-reaching cost and collision-avoidance constraints to agent's neighbors in its observation. The constraints are relaxed by slack variables to ensure the feasibility of the problem. We reported the best performance from different planning horizons. %Note that the MPC method is very time-consuming during the inference stage because it needs to solve an optimization problem at every time step for each of the agents.
% \input{tables/table1}

%Note that the MPC method is very time-consuming during the inference stage because it needs to solve an optimization problem at every time step for each of the agents.

% for each agent with a goal-reaching cost and collision-avoidance constraints. The constraints are relaxed by slack variables to ensure the feasibility of the problem. We reported the best performance from different planning horizons. 
% \textbf{RAIL:} RAIL stands for reward-augmented imitation learning, which introduces a handcrafted reward related to multiple safety criteria (distance to neighbours/road edges and hard brake frequency). Currently it achieves the closest performance (collision rate, off-road duration and hard brake rate) to the expert policy on NGSIM.

\begin{figure}[!htbp]
\begin{subfigure}[b]{0.24\textwidth}
\includegraphics[width=1.0\textwidth]{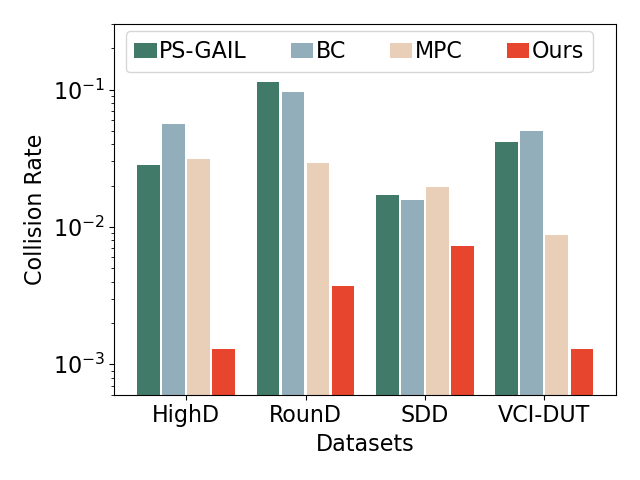}
\centering
\caption{Collision rate. %(ratio of the states in the unsafe set. Smaller is better). 
(Lower is better. The y-axis is in log scale)}
\end{subfigure}
\begin{subfigure}[b]{0.24\textwidth}
\includegraphics[width=1.0\textwidth]{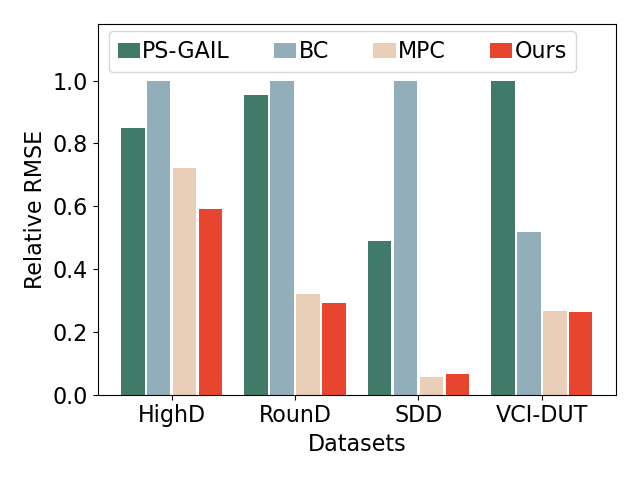}
\centering
\caption{Relative RMSE comparing to expert policy. (Lower is better)}
\end{subfigure}
\caption{Results on HighD, RounD, SDD and VCI-DUT.}
\label{fig:four}
\end{figure}

\noindent \textit{\underline{Metrics:}}
We compute the following measurements for safety and similarity to real-world data:
% measures to show not only can our approach ensure the system safety, but it also implicitly mimics experts' behaviors from real world data:
\textbf{Collision rate} reflects systems' safety, and is calculated by the number of agent states that fall into the unsafe set divided by the total number of agent states. 
\textbf{Root Mean Square Error (RMSE)} reflects how much the agents' policy diverges from the expert demonstrations. This is computed by the root mean square error of the position comparing to the ground-truth position shown in expert trajectories. For HighD, RounD, SDD and VCI-DUT, we present the relative RMSE which is further normalized under each dataset by the highest RMSE from all approaches.
\textbf{Driving related metrics:} For NGSIM, we also report other metrics related to driver behaviors on highways, such as off-road rates, hard-brake rates and average lane changes per agent as discussed in PS-GAIL and RAIL.

\subsection{Main results}
\nitb{Results on HighD, RounD, SDD and VCI-DUT:} \figref{fig:four} shows the collision rate and RMSE comparison with BC, PS-GAIL, and MPC. Our approach can achieve $53.8\%\sim 97.7\%$ reduction in collision rates across all datasets. We can see that model-free methods (PS-GAIL and BC) only get similar collision rates as MPC under simple scenarios such as HighD (where agents mostly move along the same direction with constant speeds) and SDD (large free space with low-speed pedestrians and cyclists). Though MPC is quite competitive as it directly optimizes under collision-avoidance constraints, it needs to solve an online optimization problem with complex constraints for each agent. For each simulation step, it takes 0.5 second for the MPC solver (therefore cannot really be used online), whereas our approach only needs 0.01 second, being 50X faster. 
Note that our approach still could not achieve zero collisions, which is probably caused by: 1) noisy measurements from the expert trajectories (especially, SDD is a very noisy dataset and therefore our method has a higher collision rate on SDD than other scenarios) 2) the learned CBF is only valid on training samples, and there is a  distribution shift between the training and testing data.

Surprisingly, our approach can also achieve much lower RMSE comparing to imitation learning. This implies that fixing the initial and destination, by moving safely we get simulation models that are even closer to human behaviors. Of course the high-level decision contributes to reducing RMSEs. Again only on the HighD dataset can IL methods achieve a smaller RMSE, which shows their limitations for handling complex scenarios. Using the same high-level policy and reference controllers, we can also have smaller RMSE than the model-based approach MPC. This presents the great potential of combining model-based approaches and learning frameworks, which can produce both safer and more realistic models for human road users.
% , which could lead to a promising research direction for traffic simulation and safe planning.

% \begin{table*}[!tbp]
% \begin{center}
% \footnotesize
% \begin{tabular}{cccccccccc}
% \toprule
% % && \multicolumn{7}{c}{Models}\\
% % \cline{3-9}
% \def\arraystretch{1.0}
% Metrics & Dataset & Ours & IDM\textsubscript{$\theta$}\cite{bhattacharyya2020online} & IDM\cite{treiber2017intelligent} & GAIL \cite{bhattacharyya2018multi} & Const. Speed & Const. Acc & Non-Linear Fit\cite{morton2016analysis} \\
% \hline
% \multirow{2}{*}{\makecell{Position\\RMSE}}
% & NGSIM & $2.56\pm0.65$ & $5.90\pm 1.98$ & $27.78\pm5.40$& $10.42\pm3.73$ &$6.24\pm2.02$ & $12.64\pm 4.70$ & $7.34\pm 4.55$\\
% & HighD &$1.49\pm 0.81$ & $8.02\pm 3.34$ & $18.30\pm9.03$ & $13.63\pm 3.92$ & $2.42\pm 1.64$ & $11.01\pm1.92$ & $35.13\pm7.21$\\
% \midrule
% \multirow{2}{*}{\makecell{Velocity\\RMSE}} 
% & NGSIM & $1.37\pm0.45$ & $2.12\pm 0.79$ & $10.72\pm 2.36$ & $3.52\pm 1.28$ & $2.22\pm0.82$ & $5.03\pm 1.78$ & $2.69\pm 1.77$\\
% & HighD & $0.82\pm0.50$ & $2.14\pm 0.65$ & $4.59\pm2.46$ & $2.94\pm0.93$ & $0.94\pm0.57$ & $4.39\pm0.61$ & $10.05\pm2.07$\\
% \midrule
% \multirow{2}{*}{\makecell{Number of\\collisions}} 
% & NGSIM & $0\pm 0$ & $0\pm 0$ & $0\pm 0$ & $53\pm 11$ & $113\pm18$ & $119\pm16$ & $0\pm 0$ \\
% & HighD & $0\pm0$& $0\pm 0$ & $0\pm 0$ & $15\pm 4$ & $0\pm 0$ & $27\pm3$ & $0\pm 0$\\
% \bottomrule
% \end{tabular}
% \end{center}
% \caption{Results on NGSIM and HighD under low-density scenarios (20 agents, 5 seconds).}
% \label{tab:2}
% \end{table*}

\begin{table}[!tbp]
\setlength{\tabcolsep}{1.0pt}
\begin{center}
\footnotesize
\begin{tabular}{ccccccc}
\toprule
% && \multicolumn{7}{c}{Models}\\
% \cline{3-9}
\def\arraystretch{1.0}
Metrics & Dataset & Ours & IDM\textsubscript{$\theta$}\cite{bhattacharyya2020online} & IDM\cite{treiber2017intelligent} & GAIL \cite{bhattacharyya2018multi} \\
\hline
\multirow{2}{*}{\makecell{Position\\RMSE}}
& NGSIM & $2.56\pm0.65$ & $5.90\pm 1.98$ & $27.78\pm5.40$& $10.42\pm3.73$ \\
& HighD &$1.49\pm 0.81$ & $8.02\pm 3.34$ & $18.30\pm9.03$ & $13.63\pm 3.92$ \\
\midrule
\multirow{2}{*}{\makecell{Velocity\\RMSE}} 
& NGSIM & $1.37\pm0.45$ & $2.12\pm 0.79$ & $10.72\pm 2.36$ & $3.52\pm 1.28$ & \\
& HighD & $0.82\pm0.50$ & $2.14\pm 0.65$ & $4.59\pm2.46$ & $2.94\pm0.93$ & \\
\midrule
\multirow{2}{*}{\makecell{Number of\\collisions}} 
& NGSIM & $0\pm 0$ & $0\pm 0$ & $0\pm 0$ & $53\pm 11$ \\
& HighD & $0\pm0$& $0\pm 0$ & $0\pm 0$ & $15\pm 4$\\
\bottomrule
\end{tabular}
\end{center}
\caption{Results on NGSIM and HighD under low-density scenarios (20 agents, 5 seconds).}
\label{tab:2}
\end{table}

\begin{figure}[!htbp]
\begin{subfigure}[b]{0.24\textwidth}
\includegraphics[width=1.0\textwidth]{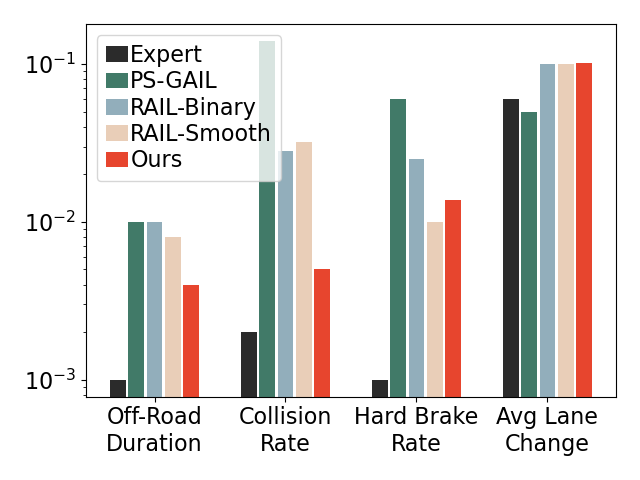}
\centering
\caption{\footnotesize Comparison for driver-related metrics with the expert policy, PS-GAIL and RAIL using the same training/testing splits as in~\cite{bhattacharyya2018multi} and~\cite{bhattacharyya2019simulating}. 
% \chuchu{
The y-axis is in logarithm scale.
%}
}
\end{subfigure}
\begin{subfigure}[b]{0.24\textwidth}
\includegraphics[width=1.0\textwidth]{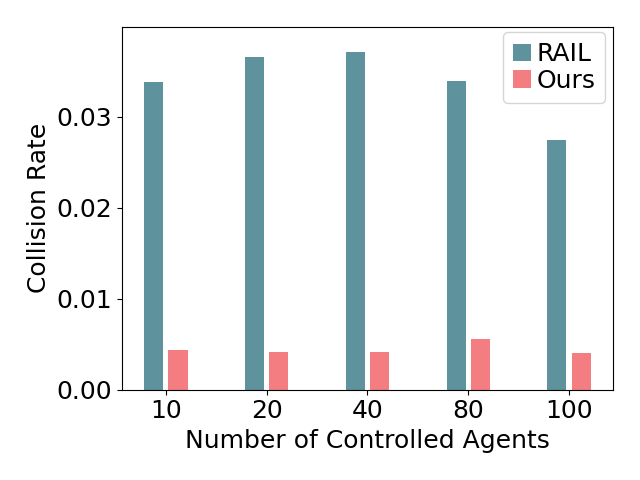}
\caption{\footnotesize Comparison of generalizability: Comparing to RAIL, our method can still maintain a low collision rate when testing traffic density is different from the one used in training.}
\end{subfigure}
\caption{Comparisons on the NGSIM dataset.}
\label{fig:ngsim}
\end{figure}

% \begin{figure*}[htb]
% \begin{subfigure}[b]{0.18\textwidth}
% \includegraphics[width=0.99\textwidth]{figs/cbf/00000_expert_VCI_lite.png}
% \caption{Screenshot for VCI.}
% \end{subfigure}
% \begin{subfigure}[b]{0.21\textwidth}
% \includegraphics[width=0.99\textwidth]{figs/cbf/VCI_tmp_0_new_heat.png}
% \caption{CBF heatmap on VCI.}
% \end{subfigure}
% \begin{subfigure}[b]{0.30\textwidth}
% %\includegraphics[width=0.99\textwidth]{figs/cbf/07705.png}
% \includegraphics[width=0.99\textwidth]{figs/cbf/RounD_00000_expert_bg1_lite.png}
% \caption{Screenshot for RounD.}
% \end{subfigure}
% \begin{subfigure}[b]{0.31\textwidth}
% \includegraphics[width=0.99\textwidth]{figs/cbf/RounD_tmp_0_new_heat.png}
% \caption{CBF heatmap on RounD.}
% \end{subfigure}

% \centering
% \caption{CBF visualization on VCI/RounD datasets.}
% \label{fig:cbf-vis}
% \end{figure*}

\nitb{Results on NGSIM and HighD:} We first test our approach for NGSIM and HighD datasets under low density traffic (20 agents, no non-reactive agents, 5 seconds) as what was used in~\cite{bhattacharyya2020online}. We compare with $\text{IDM}_\theta$~\cite{bhattacharyya2020online}, PS-GAIL~\cite{bhattacharyya2018multi} and a classical car-following model IDM~\cite{treiber2017intelligent}. As shown in Table.\ref{tab:2}, without any calibration, our approach can achieve zero collision rates on both datasets and obtains the lowest RMSE in position and velocity, which shows the power of our learned CBF in terms of providing safety guarantees.
% , which again shows the benefits of combining model-based and learning approaches.  
For challenging dense traffic cases in NGSIM (100 agents together with around 70 other non-reactive agents following expert trajectories, 20 seconds), we further compare with PS-GAIL~\cite{bhattacharyya2018multi} and RAIL~\cite{bhattacharyya2019simulating} as shown in \figref{fig:ngsim}. We also put the ``expert policy" in the figure for reference. Here ``RAIL-Binary" adopted a binary augmented reward, whereas ``RAIL-Smooth" introduced a continuous reward. %Ideally, the number of lane-changings per agent should be on a par with the statistics as the ``expert", while other metrics such as off-road rates, collision rates and hard-brake rates should reach zero. 
With similar lane changing rates\footnote{We calculate lane change rates as the number of lane-changings in the 20-second-simulation divided by the total number of agents, which is different than in \cite{bhattacharyya2019simulating}. We believe our calculation is more accurate, as supported by the statistics shown in~\cite[Figure~3.2.]{lee2004comprehensive}. RAIL's lane changing rates are updated using our new calculation method.}, our collision rate is almost $1/10$ to the ``RAIL-Smooth". Our method also achieves a lower off-road duration. Comparing to RAIL, our approach does result in a slightly higher hard brake rate, which might due to the collision-avoidance intention in our CBF controller. Our RMSE is similar to RAIL and thus is not reported here.

To show the generalizability of our approach, we train the vehicle in NGSIM using only $20$ agents then test on an increasing number of agents from $10$ to $100$. Our learned model can adapt to different traffic density and maintain a similar level of collision rates as shown in~\figref{fig:ngsim} (b). As a comparison, RAIL is trained using $100$ agents but behaves poorly when the density changes. When the density is low ($N\leq 40$),  the collision rate for RAIL goes even higher.

\subsection{Planning for ego-vehicles with reactive and safe agents}
\begin{figure}[!htbp]
\begin{subfigure}[b]{0.5\textwidth}
\includegraphics[width=0.32\textwidth]{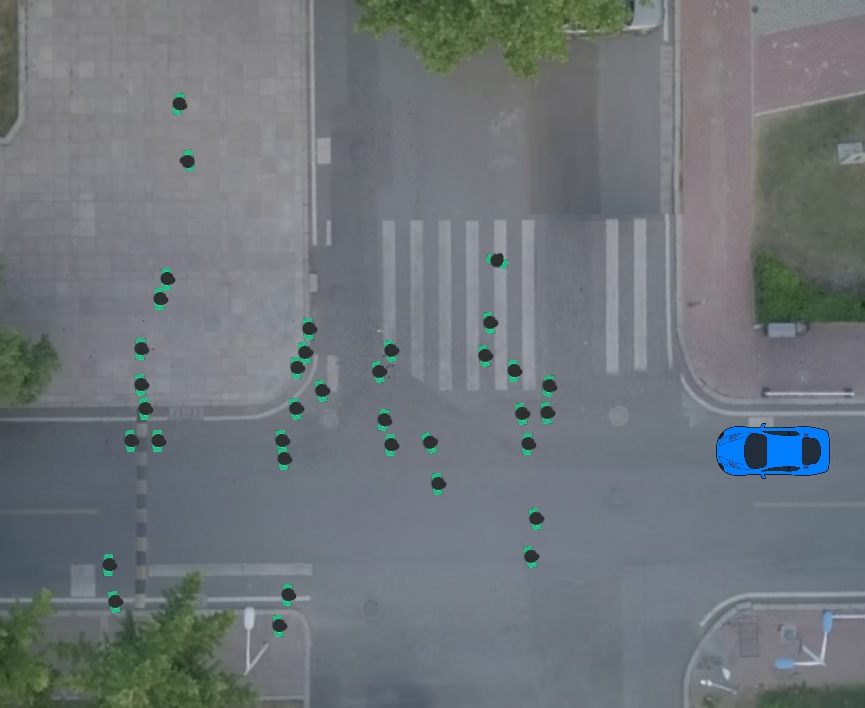} \hfill
\includegraphics[width=0.32\textwidth]{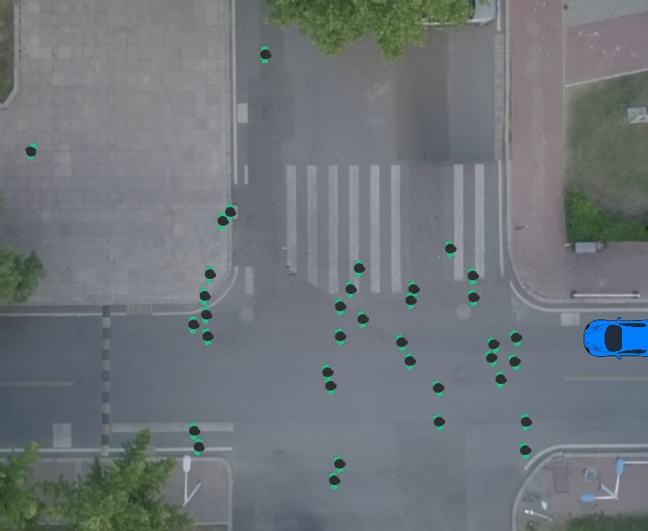} \hfill
\includegraphics[width=0.32\textwidth]{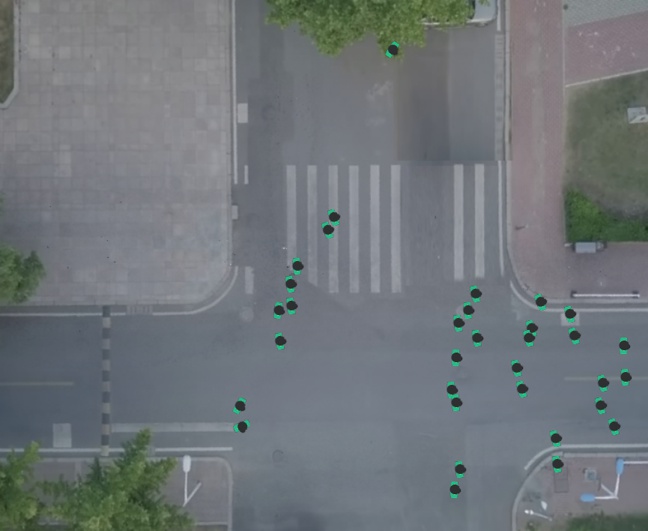}
\caption{Pedestrians that follow expert trajectories.}
\end{subfigure}
\\\\
\begin{subfigure}[b]{0.5\textwidth}
\includegraphics[width=0.32\textwidth]{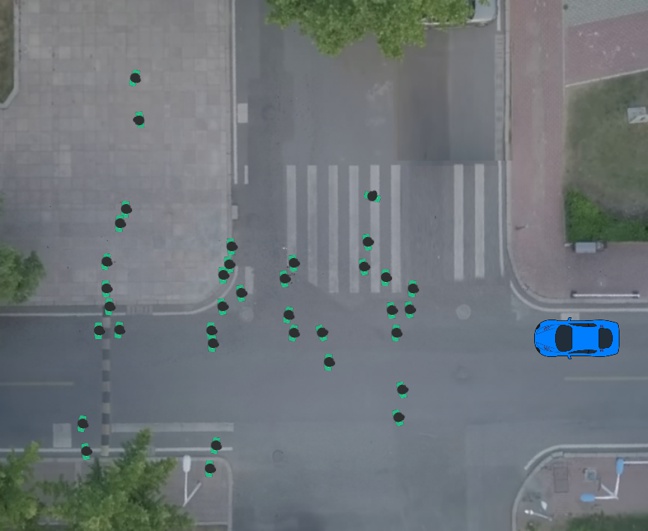} \hfill
\includegraphics[width=0.32\textwidth]{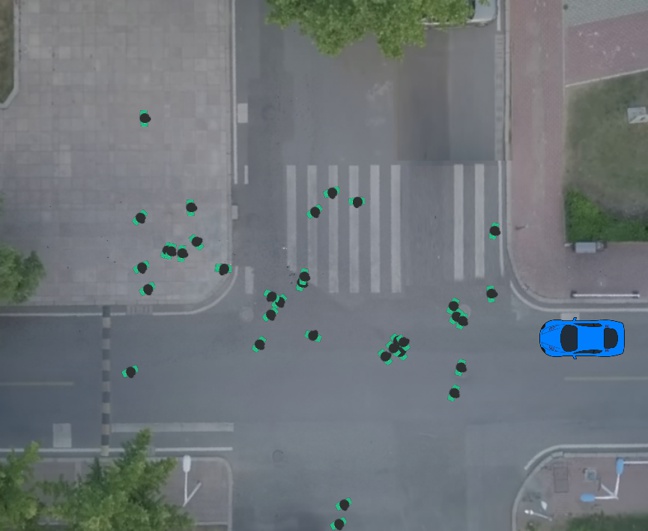} \hfill
\includegraphics[width=0.32\textwidth]{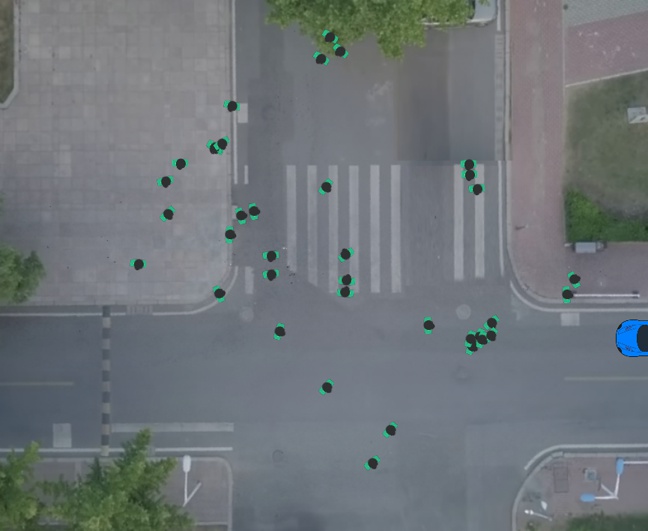}
\caption{Pedestrians running PS-GAIL models.}
\end{subfigure}
\\\\
\begin{subfigure}[b]{0.5\textwidth}
\includegraphics[width=0.32\textwidth]{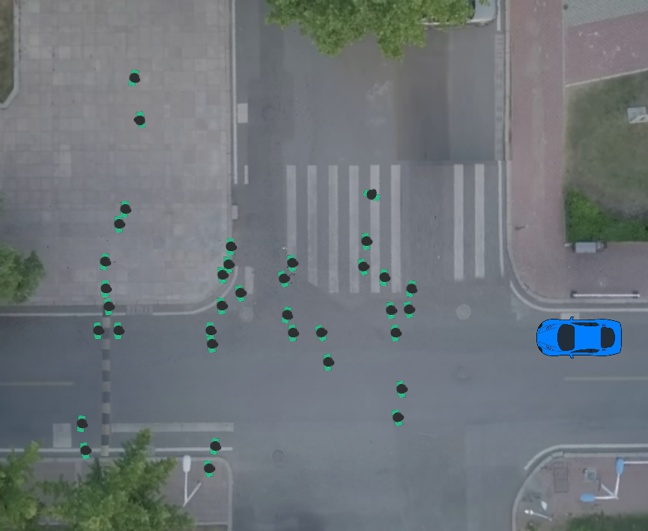} \hfill
\includegraphics[width=0.32\textwidth]{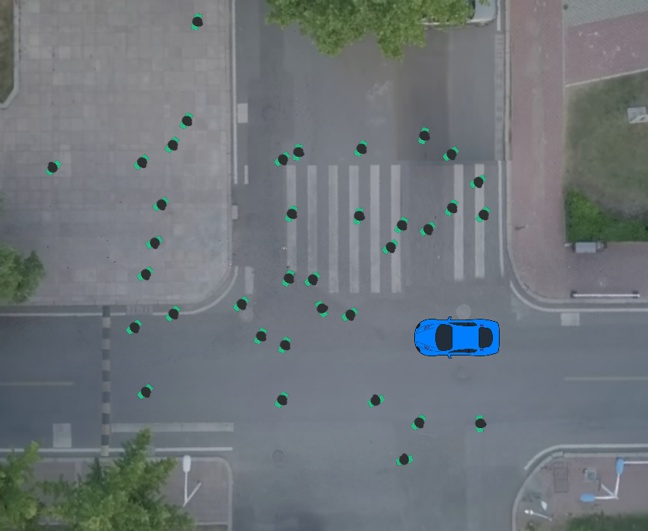} \hfill
\includegraphics[width=0.32\textwidth]{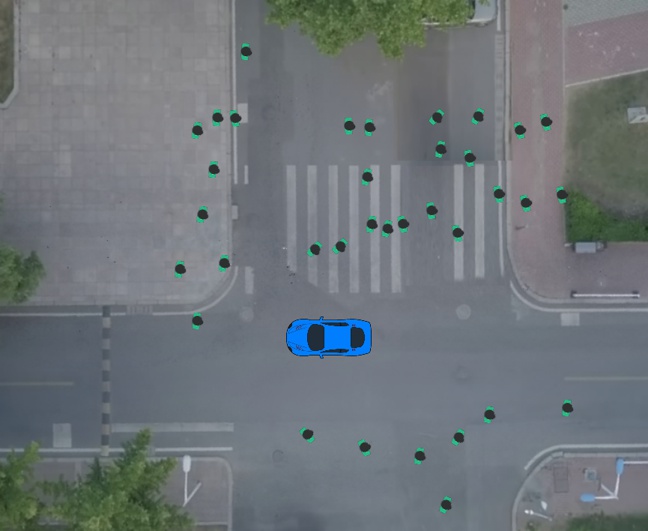}
\caption{Our learned reactive and safe pedestrians. }
\end{subfigure}
\centering
\caption{Simulations for a vehicle attempting to drive through a crowd with different pedestrian models.}
\label{fig:mpc}
\end{figure}
%  Agents following expert trajectories (a) or controlled by PS-GAIL (b) will make the ego vehicle fail to drive forward, whereas the reactive agents (c) can coordinate with the ego vehicle thus the vehicle can successfully move to the left.
To demonstrate the use of reactive and safe agents we have learned, we design a challenging planning task for a vehicle to drive through a crowded street, where the scene is taken from VCI-DUT. We consider three different pedestrian models: 1) agents that follow fixed expert trajectories 2) agents built using PS-GAIL and 3) our learned reactive agents. We use the same MPC controller for the car to drive through the crowd while enforcing safe distances to pedestrians. As shown in \figref{fig:mpc}, when interacting with our learned reactive agents, the MPC-controlled car can successfully travel through the crowd safely, as our learned agents use a neural-CBF to produce safe reactions and therefore make a path for the car. Whereas in other cases the car is frozen and even goes backward as the pedestrians are not reacting to make a way for the car. This shows safety-guided reactive agents can help solve some of the conservative issues in the developing self-driving algorithms. Note that in reality whether we can assume the pedestrians will always keep safe is another question that is out of the scope of our paper.

% To demonstrate the value of reactive agents guided by our CBF controller, we designed a challenging task for a vehicle to drive through a crowded street, where the scene was taken from VCI-DUT. We considered three different pedestrian models: 1) agents that follow fixed expert trajectories 2) agents that are controlled by PS-GAIL and 3) reactive agents that are controlled by our approach. We used model predictive control to help the vehicle drive through the crowd while enforcing a large penalty on car-pedestrian collisions. As shown in \figref{fig:mpc}, when interacting with reactive agents, since the agents learned how to avoid collisions with the coming vehicle, they left enough space for the vehicle to successfully plan for a path without collisions. Whereas the agents controlled by expert trajectories or PS-GAIL didn't react to the vehicle and due to the expensive cost in crashing, the vehicle became "conservative" and could not move forward to reach the goal. This illustrates a great potential value of our proposed learning framework for realistic simulation and for planning problems.

% \begin{figure}[htb]
% \includegraphics[width=0.5\textwidth]{figs/ngsim-exp.png}
% \centering
% \caption{NGSIM result}
% \end{figure}

%For NGSIM and HighD datasets, the reference controller is for lane keeping/lane changing. For VCI and SDD datasets, the reference controller is to let the agents move towards the destination points. For the RoundD dataset, the reference controller is to guide the agents follow trajectories to enter the roundabout and exit the roundabout.
\section{Conclusion} 
\label{sec:conclusion}

In this paper we proposed a way to generate reactive and safe road participants for traffic simulation. We start from a safety-guided learning approach, achieve the best performance among several real world traffic datasets and show the advantage of reactive agents in planning. Future work includes: 1) combine our CBF framework with imitation learning to further guide agents to behave like human and 2) integrate our models in more mature simulators like CARLA. %\chuchucomment{Also run the model in more mature simulators.}

%In this paper we proposed a way to do hybrid cbf for imitation learning to give safety certificate. We show the improvement in multiple datasets and shows it generalizabity
\section*{Acknowledgment} %ACKNOWLEDGMENT
%The NASA University Leadership initiative (grant \#80NSSC20M0163) provided funds to assist the authors with their research, but this article solely reflects the opinions and conclusions of its authors and not any NASA entity.

The authors acknowledge support from the NASA University Leadership initiative (grant \#80NSSC20M0163) and from the Defense Science and Technology Agency in Singapore. The views, opinions, and/or findings expressed are those of the authors and should not be interpreted as representing the official views or policies of any NASA entity, DSTA Singapore, or the Singapore Government.

%% Use plainnat to work nicely with natbib. 

\iftoggle{IROS}{
%%% IROS %%%
\bibliographystyle{IEEEtran}
\bibliography{IEEEabrv,z7_references}
}
{
%%% RSS %%%
\bibliographystyle{plainnat}
\bibliography{z7_references}
}

\end{document}